\documentclass[conference]{IEEEtran}
\IEEEoverridecommandlockouts
\usepackage{caption}

\usepackage{cite}
\usepackage{amsmath,amssymb,amsfonts,amsthm}
\usepackage{algorithmic}
\usepackage{graphicx}
\usepackage{svg}
\usepackage{textcomp}
\usepackage{xcolor}
\usepackage{comment}
\usepackage{booktabs} 
\usepackage{multirow} 
\usepackage{xurl}
\usepackage{bm}
\usepackage{acro}
\usepackage{tikz}
\usetikzlibrary{shapes.geometric, arrows, positioning, fit}
\usepackage[dvipsnames]{xcolor}

\DeclareAcronym{ODD}{
    short = ODD, 
    long = operational design domain
}

\DeclareAcronym{ADS}{
    short = ADS, 
    long = automated driving system
}

\DeclareAcronym{WTTC}{
    short = WTTC, 
    long = Worst-Time-To-Collision
}

\DeclareAcronym{TTC}{
    short = TTC, 
    long = Time-To-Collision
}
 
\DeclareAcronym{HiL}{
    short = HiL, 
    long = hardware-in-the-loop
}

\DeclareAcronym{SiL}{
    short = SiL, 
    long = software-in-the-loop
}

\DeclareAcronym{MiL}{
    short = MiL, 
    long = model-in-the-loop
}

\DeclareAcronym{XiL}{
    short = XiL, 
    long = x-in-the-loop
}

\DeclareAcronym{SuT}{
    short = SuT, 
    long = system under test
}

\DeclareAcronym{PG}{
    short = PG, 
    long = Proving Ground
}

\DeclareAcronym{FOT}{
    short = FOT, 
    long = Field Operational Test
}

\DeclareAcronym{ECU}{
    short = ECU, 
    long = electronic control unit
}

\DeclareAcronym{ACC}{
    short = ACC, 
    long = adaptive cruise control
}

\DeclareAcronym{AEBS}{
    short = AEBS, 
    long = advanced emergency braking system
}

\DeclareAcronym{HAD}{
    short = HAD, 
    long = highly automated driving
}

\DeclareAcronym{ADAS}{
    short = ADAS, 
    long = advanced driver-assistance systems
}

\DeclareAcronym{ML}{
    short = ML, 
    long = machine learning
}

\DeclareAcronym{LKA}{
    short = LKA,
    long = lane keeping assist
}

\DeclareAcronym{ASIL}{
    short = ASIL,
    long = automotive safety integrity level
}

\DeclareAcronym{AV}{
    short = AV,
    long = automated vehicle
}

\DeclareAcronym{SAE}{
    short = SAE,
    long = Society of Automotive Engineers
}

\DeclareAcronym{IMU}{
    short = IMU,
    long = inertial measurement unit
}

\DeclareAcronym{ssim}{
    short = SSIM,
    long = Structural Similarity
}

\DeclareAcronym{lpips}{
    short = LPIPS,
    long = Learned Perceptual Image Patch Similarity
}

\DeclareAcronym{psnr}{
    short = PSNR,
    long = Peak Signal to Noise Ratio
}

\DeclareAcronym{mse}{
    short = MSE,
    long = Mean Squared Error
}

\DeclareAcronym{fid}{
    short = FID,
    long = Fr\'echet Inception Distance
}

\DeclareAcronym{mi}{
    short = MI,
    long = Mutual Information
}

\DeclareAcronym{ncc}{
    short = NCC,
    long = Normalized Cross-Correlation
}

\DeclareAcronym{vqgan}{
    short = VQGAN,
    long = Vector Quantized Generative Adversarial Networks
}

\usepackage[hidelinks]{hyperref}
\def\BibTeX{{\rm B\kern-.05em{\sc i\kern-.025em b}\kern-.08em
    T\kern-.1667em\lower.7ex\hbox{E}\kern-.125emX}}

\PassOptionsToPackage{svgnames}{xcolor}
\usepackage{tcolorbox}
\tcbuselibrary{skins,breakable}
\usetikzlibrary{shadings,shadows}
    {\endtcolorbox}

\begin{document}
\UseRawInputEncoding
\title{LLM-Based Meta-Configuration for a Domain-Specific Language Enabling Scalable Multimodal Data Collection}
\title{A Domain-Specific Language for LLM-Driven Trigger Generation in Multimodal Data Collection}
\author{\IEEEauthorblockN{Philipp Reis, Philipp Rigoll, Martin Zehetner, Jacqueline Henle, Stefan Otten and Eric Sax}
\IEEEauthorblockA{
FZI Research Center for Information Technology, Karlsruhe, Germany\\
Email: \{reis, philipp.rigoll, zehetner, henle, otten, sax\}@fzi.de}

}
\theoremstyle{definition}
\newtheorem{definition}{Definition}[section]
\newcommand{\yes}{\Large\color{green}{\checkmark}}      
\newcommand{\no}{\Large\color{red}{\pmb{\times}}}        
\newcommand{\neutral}{\Large\color{gray}{\textbf{--}}}   

\maketitle
\bibliographystyle{IEEEtran}
\begin{abstract}
Data-driven systems depend on task-relevant data, yet data collection pipelines remain passive and indiscriminate. Continuous logging of multimodal sensor streams incurs high storage costs and captures irrelevant data.
This paper proposes a declarative framework for intent-driven, on-device data collection that enables selective collection of multimodal sensor data based on high-level user requests.
The framework combines natural language interaction with a formally specified domain-specific language (DSL). Large language models  translate user-defined requirements into verifiable and composable DSL programs that define conditional triggers across heterogeneous sensors, including cameras, LiDAR, and system telemetry.
Empirical evaluation on vehicular and robotic perception tasks shows that the DSL-based approach achieves higher generation consistency and lower execution latency than unconstrained code generation while maintaining comparable detection performance. The structured abstraction supports modular trigger composition and concurrent deployment on resource-constrained edge platforms.
This approach replaces passive logging with a verifiable, intent-driven mechanism for multimodal data collection in real-time systems.
\end{abstract}

\bstctlcite{BSTcontrol}

\section{Introduction}
\begin{table*}[t]
\centering
\caption{Comparison of Triggering and Data Collection Frameworks.}
\label{tab:trigger_framework_comparison}
\begin{tabular}{lcccc}
\toprule
\textbf{Work} & Modulare Framework & Multimodality   & Declarative & Online Selection\\
\midrule
SOLA \cite{RigollRies2024}  & $\yes$      & $\no$        & $\yes$        & $\no$         \\
Uniqueness \cite{RiesUniqueness}  & $\yes$      & $\no$        & $\no$        & $\yes$          \\
CERN \cite{collaboration_cms_2017}  & $\yes $      & $\no$        & $\no$        & $\yes$          \\
ScenarioRetrieval \cite{Sohn24vlm}  & $\yes $      & $\no$        & $\yes$        & $\no$          \\
ViperGPT \cite{viperGPT23}  & $\no $      & $\no$        & $\yes$        & $\no$           \\
RegEx \cite{Elspas22}  & $\yes $      & $\no$        & $\no$        & $\yes$       \\
McityEngin \cite{bogdoll2025mcitydataengineiterative}  & $\no $      & $\no$        & $\yes$        & $\no$         \\
DoraemonGPT\cite{yang2025doraemongptunderstandingdynamicscenes} & $\no $      & $\no$        & $\yes$        & $\no$         \\
\midrule
\textbf{Ours (LLM + DSL)}  & $\yes$      & $\yes$       & $\yes$       & $\yes$         \\
\bottomrule
\end{tabular}
\end{table*}

In data-driven development, the quality of the data fundamentally determines the reliability and performance of the resulting system~\cite{Bach17}. Vehicles with an autonomous driving sensor setup generate vast amounts of multimodal data streams. Collecting all data is economically and operationally prohibitive. Transmission and storage costs scale linearly with volume, post hoc filtering introduces latency into decision-making, and downstream teams are left curating large quantities of irrelevant samples before any learning can begin. As a result, data collection must shift from indiscriminate logging to selective capture that prioritizes samples most informative for the intended downstream tasks.
\begin{figure}
    \centering
    \includegraphics[width=\linewidth]{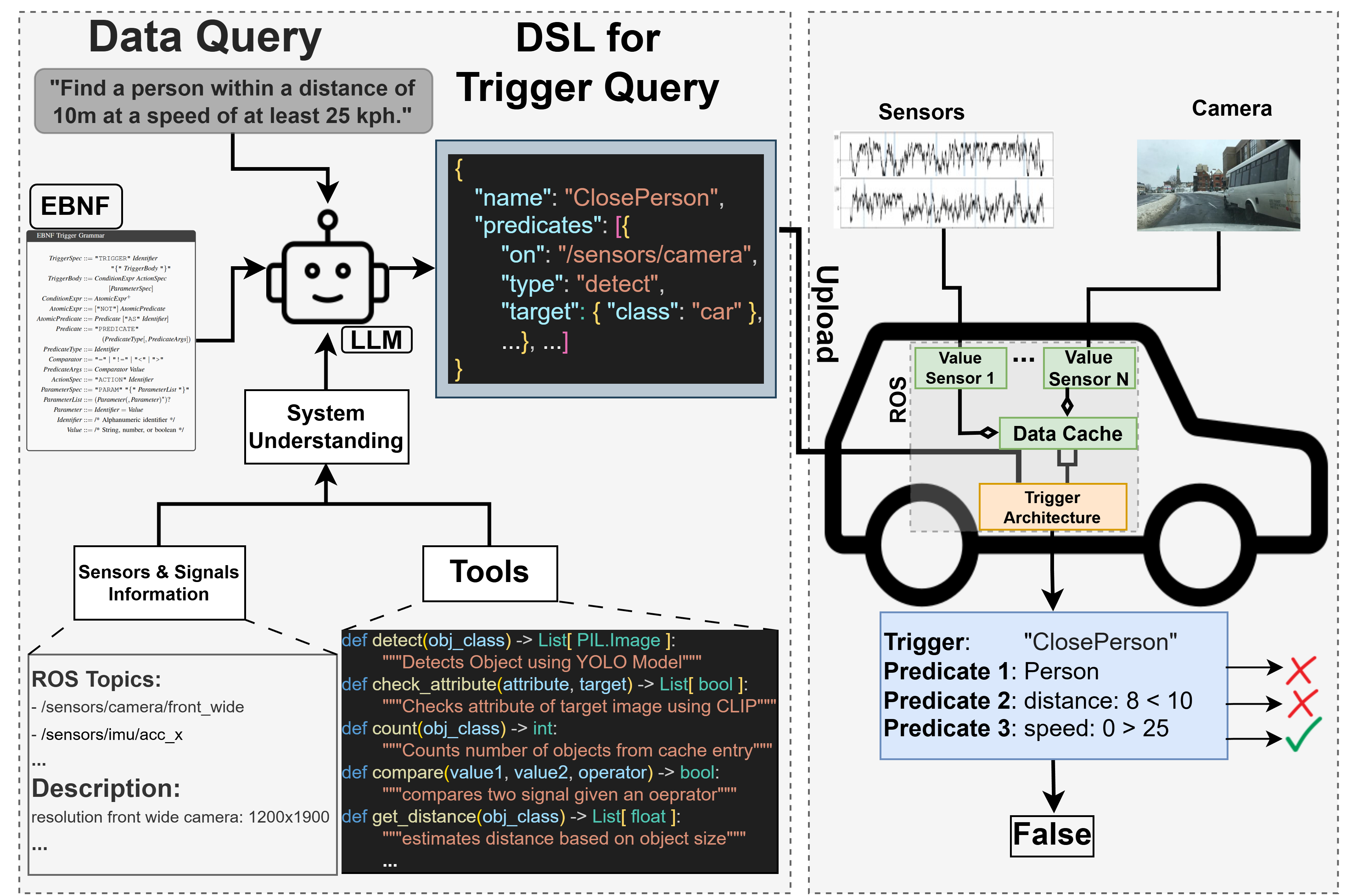}
    \caption{Proposed framework for DSL-based trigger generation. Given a data query, available tools, and system information, the framework generates a structured trigger configuration, which is deployed to the vehicle trigger architecture for selective data collection.}
    \label{fig:framework}
\end{figure}

Existing approaches to selective capture data typically focus on corner cases or anomalies on different modalities and abstraction levels, ranging from pixel-level to scenario-level \cite{Breitenstein_20,bogdoll_anomaly_2022}. While effective in narrow settings, these strategies exhibit three recurring limitations. First, they depend on expert-crafted, low-level code that is difficult to audit, reuse, and maintain across platforms and sensor configurations. Second, they often compose multiple, independently implemented triggers whose parallel execution is inefficient and whose interactions are undefined, leading to missed events or duplicated captures. Third, they equate “relevance” with generic anomaly signals; however, relevance is user- and task-dependent: what is relevant for perception robustness  may differ from what matters for control or driving function evaluation. Consequently, current pipelines lack a comprehensive and scalable method to express, verify, and deploy data relevance policies that adapt to evolving objectives.

This paper reframes selective data collection as a declarative specification problem: stakeholders should describe what to collect in natural language, and the system should ensure efficient and on-device trigger execution and recording over heterogeneous sensors, see \autoref{fig:framework}.

The main contributions of this paper are:

\begin{enumerate}
    \item  An \textbf{Domain-Specific Language} (DSL)  for multimodal data triggers with conditional logic, and composition, enabling formal verification and efficient execution on the device.
    \item A \textbf{multimodal trigger framework} that leverages large language models to translate user requests into auditable, reusable capture programs, reducing the need for specialized system and code expertise.
    \item An \textbf{empirical evaluation} on vehicular and robotic platforms demonstrating improved trigger efficiency, reduced redundant capture, and enhanced dataset quality for training and evaluation.
\end{enumerate}

By elevating selective data capture from hand-written heuristics to declarative, verifiable specifications, this approach aligns data collection with the needs of downstream tasks, acknowledges that relevance is user-dependent, and delivers a practical path to scalable, efficient, and robust data-centric AI at the edge.

\section{Related Work}
Data collection strategies can be categorized into dataset-specific and data-intrinsic approaches. Dataset-specific strategies aim to improve dataset quality by analyzing global properties, such as distribution, coverage, and statistics of the dataset \cite{reis2025datadrivennoveltyscorediverse, reis2025feedbackcontrolframeworkefficientdataset}.
In contrast, data-intrinsic collection strategies focus on the inherent properties of a datapoint itself \cite{heidecker21,collaboration_cms_2017}  and evaluate the relevance for a specific data need. These properties can be identified using predefined rules, semantic understanding, or model performance indicators such as performance metrics.

In order to systematically compare existing data collection strategies, we evaluate prior work along four key dimensions that reflect the requirements of practical in-vehicle deployment and user-centered trigger design.
First, \textbf{modular framework} refers to whether a method is embedded in a flexible system architecture that allows triggers to be added, removed, or modified independently. In large-scale vehicle fleets, data needs evolve continuously (e.g., new edge cases, new perception modules, new validation targets). A modular framework enables the incremental integration of new triggers without requiring redesign of the entire pipeline. Traditional rule-based strategies~\cite{Elspas22,collaboration_cms_2017} often lack such abstraction and require manual integration at system level, while more recent semantic and learning-based approaches are sometimes tightly coupled to specific perception stacks \cite{bogdoll2025mcitydataengineiterative,RiesUniqueness}.
Second, \textbf{online selection} captures whether a strategy can operate in real time under vehicle resource constraints. Data triggers in vehicle systems must execute efficiently on embedded hardware and make decisions during driving without introducing computational overhead. Rule-based triggers are efficient due to their simplicity, whereas semantic-based approaches~\cite{Sohn24vlm,RigollRies2024}  require increasing runtime cost. 
Third, \textbf{declarative} design describes whether triggers can be specified at a high level without requiring detailed system or programming knowledge. In classical rule-based systems\cite{Elspas22,collaboration_cms_2017}, triggers must be manually implemented in code and require a deep understanding of signal interfaces and software architecture. Error-based methods~\cite{Pathuri24} require expert configuration and integration. In contrast, declarative trigger design enables users to express data needs in natural language or domain-specific abstractions, lowering the barrier for non-expert stakeholders such as validation engineers or data analysts.
Fourth, \textbf{multimodality} evaluates whether a method can operate jointly on heterogeneous sensor streams such as camera images, LiDAR point clouds, radar signals, and vehicle telemetry. Many existing approaches focus primarily on vision data or operate on single modalities. For example, vision-language systems such as ViperGPT~\cite{Suris_2023_ICCV} provide a natural-language interface for visual reasoning but are limited to unimodal image or video inputs and are not designed for synchronized multi-sensor system deployment. However, practical vehicle data collection requires reasoning across multiple modalities.

These four dimensions, modularity, online capability, declarative specification, and multimodality form the basis of the evaluation criteria summarized in Table~\ref{tab:trigger_framework_comparison}. They highlight limitations of existing approaches and motivate the need for an LLM-driven framework that translates user intents into executable domain-specific triggers deployable in real-time, modular vehicle data collection systems.

\subsection{Problem Description}

Existing data collection strategies require expert knowledge to translate high-level data needs into executable trigger logic. Stakeholders such as validation engineers or safety experts typically formulate requirements at a semantic level (e.g., collect rare pedestrian interactions in urban rain scenarios), while deployment in vehicle fleets demands concrete, resource-efficient, and modular trigger implementations tightly aligned with the software stack.
This gap between stakeholder intent and deployable trigger logic currently leads to manual, time-consuming development processes. Implementing triggers requires detailed knowledge of signal interfaces, perception outputs, runtime constraints, and system integration. As a result, trigger development becomes a bottleneck and limits scalability and rapid iteration in fleet-wide data collection.

To overcome this challenge, a domain-specific language (DSL) is required that provides a structured and deployment-ready representation of data triggers. Such a DSL must formalize predefined abstractions and rules aligned with the underlying data collection framework, ensuring modular integration, real-time feasibility, and multimodal expressiveness.

\section{Domain Specific Language for Trigger Generation}
\subsection{Definitions and Problem Statement }
To enable efficient and scalable data collection in vehicle environments, we propose a formal framework for defining \textit{triggers}—conditions under which data should be recorded. This requires formalizing the key concepts involved in scene interpretation and vehicle state monitoring.

\begin{definition}[Scene $S$]
    A scene describes a specific snapshot of a scenario \cite{Ulbrich15}. Based on \cite{reis25Caps}, it consists of a finite set of elements $e_i$:
    \begin{equation}
        S = \{e_\mathrm{1},e_\mathrm{2},\dots, e_\mathrm{N}\}.
    \end{equation}
    Each element $e_i$ is defined by the spatial and semantic property $\phi$ tuple:
    \begin{equation}
        e_i = (\phi_\mathrm{sem}, \phi_\mathrm{spat}).
    \end{equation}
\end{definition}
\begin{definition}[Vehicle State $\mathcal{V}$]
    A vehicle state refers to all internal signals of the ego vehicle, its measurable internal conditions and controls as a tuple of data type, value and unit:
    \begin{equation}
        \mathcal{V} = \{v_\mathrm{1},v_\mathrm{2},..,v_\mathrm{m}\},\quad v_i :=(\text{type},\text{value},\text{unit}).
    \end{equation}
\end{definition}
\begin{definition}[Situation $\mathbb{S}$]\label{def:situation}
A vehicle situation is the set of the most recent values of a scene and the vehicle status at a specific point in time:
    \begin{equation}
        \mathbb{S} = \mathcal{S} \cup \mathcal{V}.
    \end{equation}
\end{definition}
\begin{definition}[Predicate $\mathbf{P}$]
    A predicate is a Boolean function that evaluates a specific condition on a particular value or attribute within a situation:
    \begin{equation}
        \mathbf{P}:\mathbb{S} \rightarrow \{True,False\},\quad P(\mathbb{S}) = \psi(f(\mathbb{S})) 
    \end{equation}
    where $s\in\mathbb{S}$ is a specific situation of the set of all possible situations. $f(\mathbb{S})$ extracts a specific value from this situation. $\psi: \mathcal{D} \rightarrow { \text{True}, \text{False} }$ is a Boolean condition (e.g., $\psi(v) := v > X$ for a threshold $X$). The output indicates whether a specific condition is satisfied (True) or not (False) for a given situation.
\end{definition}

\begin{definition}[Trigger $\mathbf{T}$]
    A Trigger is a Boolean function that combines a finite set of predicates ${ \mathbf{P}_1, \ldots, \mathbf{P}_n }$ using logical operators AND ($\land$) and NOT ($\lnot$).
    \begin{equation}
    \begin{split}
        \mathbf{T}&:\mathbb{S} \rightarrow \{True,False\},\\
        \mathbf{T}(s) &= \Psi\begin{pmatrix}
                                \mathbf{P}_1(S), \ldots, \mathbf{P}_n(\mathcal{S}) 
                        \end{pmatrix}
    \end{split}
    \end{equation}
    where each $\mathbf{P}$ is a predicate and $\Psi: \{ \text{True}, \text{False} \}^n \rightarrow \{ \text{True}, \text{False} \}$ is a Boolean formula defining the logical structure of the Trigger $\mathbf{T}$.
\end{definition}

Based on the definition of a Trigger $\mathbf{T}$ consisting of a logical set of one or multiple predicates, a domain-specific language is defined.

\subsection{Domain Specific Language for Scene Based Trigger Generation}
\begin{figure}[t]
\centering
\begin{tcolorbox}[title=EBNF Trigger Grammar, colback=gray!5, colframe=black!75, breakable]
\begin{align*}
\textit{TriggerSpec} &::= \texttt{"TRIGGER"} \; \textit{Identifier} \; [ \textit{SpecList} ] \\
&\qquad \qquad \; \texttt{"\{"} \; \textit{ConditionExpr} \; \texttt{"\}"} \\
\textit{SpecList} &::= \texttt{"("} \; \textit{Spec} \; ( \texttt{","} \; \textit{Spec} )^* \; \texttt{")"} \\
\textit{Spec} &::= \textit{Identifier} \; \texttt{"="} \; \textit{Value} \\
\textit{ConditionExpr} &::= \textit{AtomicExpr}^+ \\
\textit{AtomicExpr} &::= [ \texttt{"NOT"} ] \; \textit{AtomicPredicate} \\
\textit{AtomicPredicate} &::= \textit{Predicate} \; [ \texttt{"AS"} \; \textit{Identifier} ] \\[0.4em]
\textit{Predicate} &::= \texttt{"PREDICATE"} 
( \textit{PredicateType}\\&\quad \qquad  \; [ \texttt{","} \; \textit{PredicateArgs} ] ) \\
\textit{PredicateType} &::= \textit{Identifier} \\
\textit{PredicateArgs} &::= \textit{ComparatorExpr} \;|\; \textit{ArgList} \\
\textit{ComparatorExpr} &::= \textit{Comparator} \; \textit{Value} \\
\textit{ArgList} &::= \textit{Arg} \; ( \texttt{","} \; \textit{Arg} )^* \\
\textit{Arg} &::= \textit{Value} \;|\; \textit{NamedArg} \\
\textit{NamedArg} &::= \textit{Identifier} \; \texttt{"="} \; \textit{Value} \\[0.4em]
\textit{Comparator} &::= \texttt{"="} \;|\; \texttt{"!="} \;|\; \texttt{"<"} \;|\; \texttt{">"}  \\
\textit{Identifier} &::= \text{/* Alphanumeric identifier */} \\
\textit{Value} &::= \text{/* string, number, or boolean */} \\
\end{align*}
\end{tcolorbox}
\caption{Definition of the DSL using the extended Backus-Naur form (EBNF).}
\label{fig:dsl}
\end{figure}
To enable precise, extensible specification of event-based triggers for complex scene understanding tasks, we define a domain-specific language (DSL) with a formal grammar. Our DSL supports compositional logical conditions over scenes and vehicle states, facilitating modular, interpretable trigger definitions that can be easily parsed and executed in real-time systems.

We define the syntax of our domain-specific language for trigger specification using Extended Backus-Naur Form (EBNF). A trigger is introduced by the keyword TRIGGER followed by an identifier and a body enclosed in braces. The trigger body consists of a condition expression, an action specification, and an optional parameter block.
Conditions are expressed as logical formulas constructed from atomic predicates combined via AND, OR, and optional negation (NOT), respecting standard operator precedence. Atomic predicates encapsulate semantic or spatial constraints and may be optionally assigned an alias using the keyword AS.

Each predicate is introduced with the keyword PREDICATE, followed by a predicate type identifier and optional arguments comprising a comparator and a value, allowing for flexible quantitative and qualitative expressions. The action specification associates the trigger with a named action to execute upon condition satisfaction.
Parameters, if present, are key-value pairs enclosed within a PARAMETERS block, enabling customization without complicating the logical structure.
For vehicle data triggering, each predicate is linked to a code snippet that meets safety and efficiency constraints for scalable and safe deployments.


\subsection{Large Language Integration}
\begin{table*}[ht] 
\caption{Mapping from EBNF defined DSL to JSON-based representation.}
\label{tab:dsl-json-mapping}
\centering
\renewcommand{\arraystretch}{1.2}
\begin{tabular}{p{4.3cm}p{7.2cm} p{3.9cm}}
\toprule
\textbf{DSL Construct} & \textbf{JSON Representation} & \textbf{Description} \\
\midrule
\texttt{TRIGGER <id> (frequency=2) \{<body>\}} &
\texttt{\{"type": "trigger", "name": "<id>", "spec": \{"frequency": 2\}, ...\}} &
Trigger-level sampling specification (e.g., evaluation rate in Hz) \\
\midrule
\texttt{PREDICATE(detect, class="person")} &
\texttt{\{"type": "detect", "target": \{"class": "person"\}\}} &
Detect object of given class \\
\midrule
\texttt{PREDICATE(<type>, on=<x>, ...)} &
Add \texttt{"on": "<x>"} to JSON &
Predicate input reference (topic name / stored predicate result) \\
\midrule
\texttt{PREDICATE(count, on="p", operator=">", value=5)} &
\texttt{\{"type": "count", "on": "p", "comparison": \{"operator": ">", "value": 5\}\}} &
Count comparison on variable \\
\midrule
\texttt{PREDICATE(... AS x)} &
Add \texttt{"store\_as": "x"} to JSON &
Stores result as variable \\
\midrule
\texttt{NOT PREDICATE(...)} &
Add \texttt{"not": true} to JSON &
Logical negation (optional) \\
\bottomrule
\end{tabular}
\end{table*}

Although the EBNF-based DSL provides a robust and generalizable formalism for defining data triggers, manually authoring trigger specifications from natural language requirements is time-consuming and error-prone. Automating this process requires capabilities in natural language understanding, system comprehension, and contextual reasoning.
Large Language Models (LLMs) can address this gap by translating user-specified data collection requirements into valid DSL constructs. Given the EBNF grammar, LLMs can generate an intermediate JSON representation of triggers (see Table~\ref{tab:dsl-json-mapping}), which serves both as an interpretable format and as input for execution engines.
However, effective LLM integration requires access to system-specific knowledge, such as vehicle signal descriptions, control logic, and sensor mappings, which are often scattered across diverse, unstructured data sources, including K-matrices, codebases, and datasheets. To facilitate this, we employ a basic retrieval-augmented generation (RAG) setup, which enhances the model's access to relevant documents.
While full system integration and knowledge representation are ongoing research topics, the present work focuses on establishing the formal and conceptual framework for trigger specification and exploring the viability of LLM-based automation in this domain.


\section{Trigger Framework for  vehicle Systems}
\subsection{Trigger generation Process}
\begin{figure*}
    \centering
    \includegraphics[width=1\linewidth]{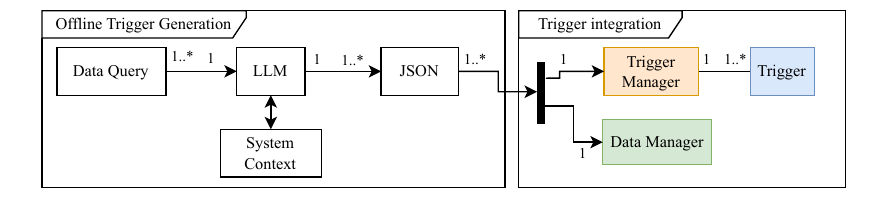}
    \caption{Workflow for scalable data collection. Data requests are processed by a LLMs to a JSON files based on the proposed DSL. Each JSON File is the configuration for an individual trigger, which is the interface for the trigger framework.}
    \label{fig:trigger_workflow}
\end{figure*}
The proposed framework enables scalable in-vehicle data collection through a structured pipeline consisting of three stages: (i) specification of data requirements, (ii) automated generation of trigger configurations, and (iii) system-level integration and configuration, see ~\autoref{fig:trigger_workflow}.
Initially, data acquisition requirements are expressed in natural language. These may originate from domain experts or be specified using structured templates, such as data cards commonly employed in large-scale data operations. Each high-level requirement is subsequently decomposed into one or more atomic predicates.
To bridge the gap between informal requirements and machine-executable configurations, a domain-specific large language model (LLM) is employed. This model, informed by both the domain-specific EBNF grammar and contextual vehicle system information, translates each data request into a structured JSON specification. The generated JSON adheres strictly to the syntax defined by the trigger grammar and represents the logical structure, parameterization, and required vehicle data inputs.
The resulting trigger configurations are forwarded to the vehicle system, where they are interpreted and managed by dedicated software components, see \autoref{fig:trigger_workflow}. These components evaluate the specified conditions in real-time and activate data capture routines if the trigger logic evaluates to true. 

\begin{figure*}
    \centering
    \includegraphics[width=1\linewidth]{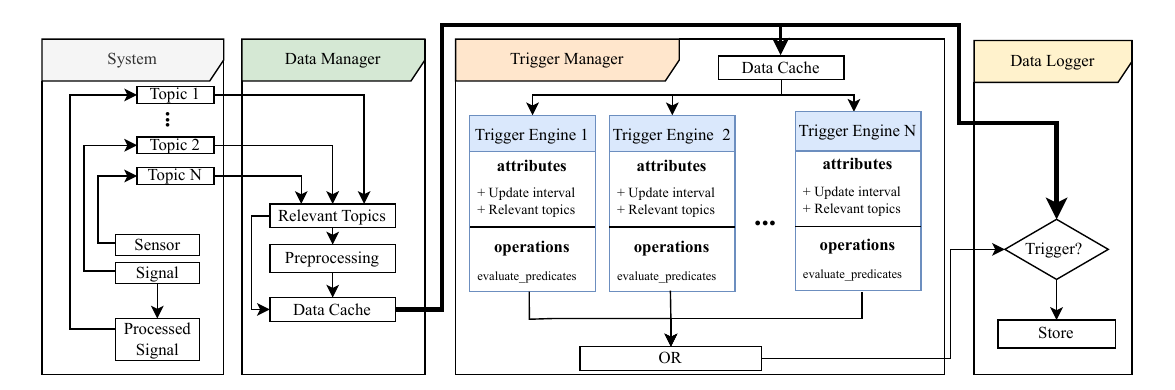}
    \caption{Data flow within the vehicle system for trigger-based data collection. Data are published to system topics, preprocessed by the Data Manager, and cached for trigger evaluation before being forwarded to the Trigger Manager. The Trigger Manager evaluates each trigger according to its configuration. Upon activation, the corresponding data cache is persisted.}
    \label{fig:data_flow}
\end{figure*}
\subsection{In-Vehicle Trigger Architecture}
The in-vehicle execution environment is structured around three principal components: the \emph{Data Manager}, the \emph{Trigger Engine}, and the \emph{Trigger Manager} within a ROS environment. Each component is responsible for a distinct phase in the real-time evaluation and execution of triggers, as shown in~\autoref{fig:data_flow}.
\subsubsection*{System}
The system reflects the connection between the hardware to the software. Sensors, signals and processed signals from distributed ECUs and sensors are communicated and provided for further processing.

\subsubsection*{Data Manager}
The Data Manager serves as the data provisioning backbone of the system. Its primary role is to supply the Trigger Engine with the subset of vehicle signals relevant for trigger evaluation. To this end, the Data Manager performs filtering, signal preprocessing, and computational optimization, ensuring that only the necessary information is extracted and cached.
A central structure maintained by the Data Manager is the \emph{data cache}, which encapsulates the current vehicle situation as defined in Definition~\ref{def:situation}. This cache is continuously updated as new sensor or vehicle bus data becomes available, and it provides a unified interface for querying both raw and derived signal states.
\subsubsection*{Trigger Engine}
Each trigger configuration generated by the LLM corresponds to a dedicated instance of the Trigger Engine. The JSON-based specification defines a finite set of predicates, their logical composition, and associated metadata such as evaluation frequency and required signal types.
At runtime, the Trigger Engine evaluates the logical condition $\mathbf{T}$ over the current vehicle situation $\mathbb{S}$. Each predicate $\mathbf{P}_i$ is associated with executable code that computes its Boolean outcome based on the data available in the cache.
The Trigger Engine operates conditionally: it evaluates the trigger only when (i) new relevant data becomes available, and (ii) the update frequency criterion is met. If the complete trigger expression evaluates to \texttt{True}, the corresponding data collection action is initiated or flagged.
\subsubsection*{Trigger Manager}
The Trigger Manager supervises all instantiated Trigger Engines. It is responsible for orchestrating their evaluation schedules, monitoring signal updates, and ensuring adherence to system-level constraints such as resource usage or priority policies.
Operationally, the Trigger Manager polls each engine to determine whether its evaluation conditions have been met. If so, it delegates execution and collects the Boolean outcome. Upon trigger activation, additional metadata (e.g., timestamp, context, affected signals) may be logged or transmitted, depending on the overarching data collection strategy.
This modular separation of responsibilities ensures that the framework remains extensible, traceable, and adaptable to varying vehicle architectures and operational scenarios.
\subsection*{Data Logger}
The output of all triggers is the input to an OR Gate, which produces a data trigger and stores the current data cache with the respective active triggers.

\begin{figure*}
    \centering
    \includegraphics[width=1\linewidth]{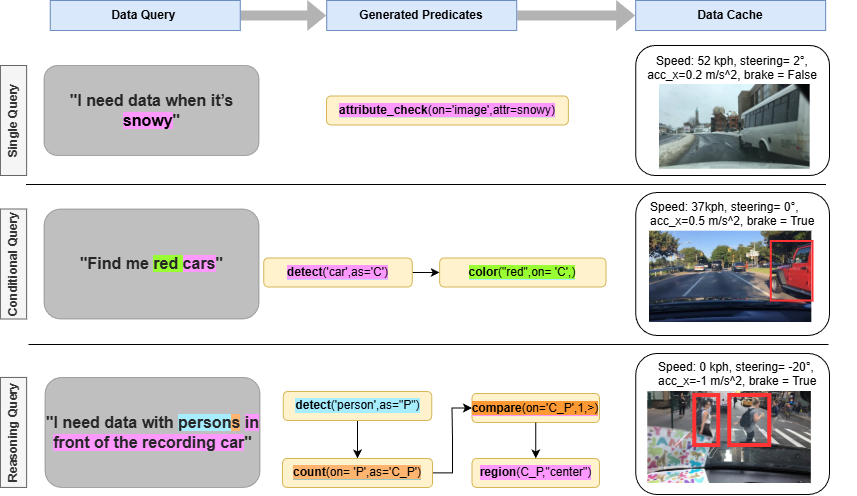}
    \caption{Three exemplary triggers generated by the proposed framework. A data query (left) is transformed into atomic predicates (middle), which are composed into a detection logic and evaluated against the current data cache (right). }
    \label{fig:examples}
\end{figure*}

Exemplary use cases of the proposed framework are illustrated in \autoref{fig:examples}.

\section{Evaluation}\label{sec:Results}

This section evaluates the proposed data-triggering framework. The assessment compares LLM-generated configurations expressed in the proposed DSL with directly generated trigger logic code. Both variants are produced using the same backbone model, gpt-oss-120b \cite{openai2025gptoss120bgptoss20bmodel}, and identical prompt templates to ensure comparability across runs.

As an additional benchmark, a vision-language model (VLM) based on gemma3:27b \cite{gemmateam2025gemma3technicalreport} is included. The evaluation examines detection capability, consistency of trigger generation, and runtime performance of the resulting data triggers.

\subsubsection{Data Query Categories}
\begin{table*}[h]
\caption{Data queries grouped by category.}
\label{tab:tasks_by_category}
\resizebox{\linewidth}{!}{%
\centering
\begin{tabular}{l l l }
\toprule
\textbf{Single Query} & \textbf{Reasoning Query} & \textbf{Conditional Query} \\
\midrule
I want to find images with stop signs. 
& I need data with persons in front of the recording car. 
& I want data with a red traffic light. \\ 

I need situations with trains. 
& I need situations with a person riding a bike. 
& I need data with more than 5 cars. \\ 

I need data when it's snowy. 
& I need data involving car-following scenarios. 
& Find me pedestrians when its snowy. \\ 

Find me situations with a minimum of 60\% brightness. 
& Find me images where the car in front is less than 5m away. 
& Find me red cars. \\ 

I want data with persons visible. 
& 
& Find me situations with a person and a car. \\ \bottomrule

\end{tabular}
}
\end{table*}

Three classes of data perception requests with increasing semantic complexity are evaluated. Single data queries request basic object detection such as finding persons. Conditional queries add spatial or attribute constraints. Reasoning queries require relational or contextual interpretation.

\subsubsection{Available Tools}
The DSL and the plain Code approach use the same perception primitives. Object detection is performed using a YOLO based detector via \texttt{run\_yolo(args)}, and attribute filtering is performed using a CLIP-based model via \texttt{run\_clip(args)}. In the DSL-based approach, further tools are implemented as predicates including \texttt{regional}, \texttt{color}, \texttt{distance}, and \texttt{value\_compare} functions.

 \subsubsection{Dataset}
A Handcrafted subset consisting of 35 images from BDD100k~\cite{Yu2018BDD100KAD}, which are manually labeled and compared against the output of the corresponding trigger system. 

\subsubsection{Evaluation Pipeline}

For each data query, the LLM generates a corresponding trigger. In the DSL and plain-code approaches, the generated trigger is executed sequentially on all images in the dataset. The resulting binary outputs are compared against the ground-truth annotations.
For the VLM approach, the data query description is provided directly into the prompt, and the model outputs a binary decision (True/False) for each image without intermediate trigger generation.
To evaluate generation consistency, each data query is generated twice using different random seeds under default temperature settings. The resulting triggers are compared to assess variability in both DSL-based and plain-code generation.

\subsection{Detection Capability}
\begin{figure}
    \centering
    \includegraphics[width=1\linewidth]{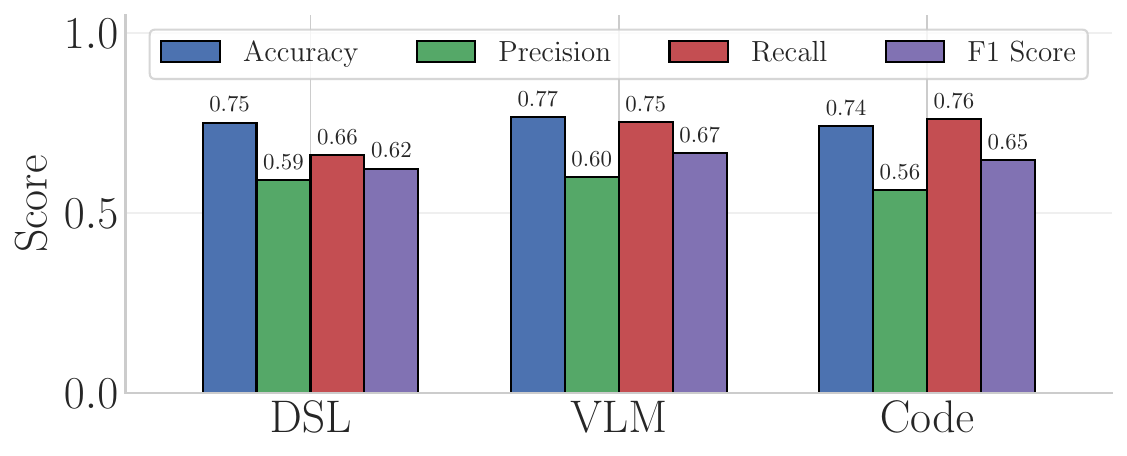}
    \caption{Comparison of detection performance for data-triggered events across the DSL framework, the code generation approach, and the VLM-based approach.}
    \label{fig:detection_metrics}
\end{figure}
Detection capability is evaluated by verifying whether an executable configuration produces a valid detection result that satisfies the requested condition. Manually annotated labels serve as ground truth for expected behavior.
Across all query types, overall performance is comparable, see \autoref{fig:detection_metrics}. The VLM achieves the highest performance with an $F_1$-score of $0.67$. The DSL-based approach shows lower recall ($0.66$) compared to the VLM ($0.75$) and the plain-code approach ($0.76$), indicating a higher rate of false negatives and reduced sensitivity in retrieving relevant samples.

\subsection{Trigger Generation Consistency}
\begin{table}[]
    \centering
    \caption{Comparision of the consistency of generated triggers between the proposed trigger framework and the plain code generation.s}
    \label{tab:consistency}
    \begin{tabular}{llll}
        \toprule
         & Sequence & Levenshtein & Cosine \\
        \midrule
        DSL & $\bm{0.966 \pm 0.067}$ & $\bm{0.955 \pm 0.106}$ & $\bm{0.918 \pm 0.111}$ \\
        Code & $0.392 \pm0.225$ & $0.510\pm0.207$ & $0.802 \pm 0.134$ \\
        \bottomrule
    \end{tabular}
\end{table}

To evaluate generation consistency of triggers produced by the DSL and plain-code approaches, three similarity metrics are applied: SequenceMatcher similarity, Levenshtein similarity, and cosine similarity.
SequenceMatcher similarity quantifies overlap by identifying the longest contiguous matching subsequences, emphasizing preserved character order and local structural similarity. Levenshtein similarity is derived from edit distance and measures the minimal number of insertions, deletions, and substitutions required to transform one code instance into another, capturing surface-level differences. Cosine similarity represents code as vector embeddings and computes the cosine of the angle between them, thereby reflecting semantic similarity even when syntactic structures differ.
Results indicate that triggers generated using the proposed DSL framework achieve higher similarity across all metrics (see \autoref{tab:consistency}). Although the plain-code approach exhibits moderate semantic similarity, substantial character-level variation is observed, indicating lower generation consistency. In contrast, DSL-generated triggers remain more consistent at both syntactic and semantic levels.

\subsection{Runtime Performance}
\begin{figure}[t]
    \centering
    \includegraphics[width=1\linewidth]{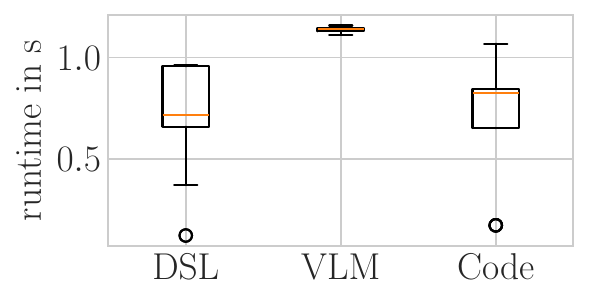}
    \caption{Execution runtime comparison between the proposed framework, the VLM-based approach, and the plain code detection approach. }
    \label{fig:runtime_comparison}
\end{figure}

End-to-end execution time is measured for each trigger configuration across all samples per data query. Runtime performance is critical in real-time systems where multiple triggers may operate concurrently. To avoid measuring no-operation behavior, only trigger configurations that produced at least one true positive detection are considered.
The proposed DSL-based framework achieves the lowest execution time, attributable to optimized backbone integration, see \autoref{fig:runtime_comparison}. In contrast, the plain-code approach frequently introduces redundant or irrelevant checks, resulting in reduced runtime efficiency.
The VLM exhibits the highest, yet most consistent, execution time, as inference cost remains independent of the specific data query.

\section{Discussion and Limitations}

The evaluation highlights several practical strengths of the proposed trigger framework for edge data collection systems. By constraining trigger specifications through a formally defined domain-specific language, the framework improves consistent trigger generation and achieves the lowest execution runtime. These properties are essential for real-time operation on resource-constrained platforms where multiple triggers may run concurrently.
Triggers generated via the DSL are inherently modular. New triggers can be added, removed, or modified independently, as adjustments affect only specific logical components. This modular design enables flexible deployment and incremental system updates without disrupting existing configurations, while preserving scalability and maintainability as data collection demands grow.
In contrast, VLM-based detection lacks scalability. Each additional trigger requires a separate inference step, and VLM inference incurs the highest computational cost per query. While the plain-code approach demonstrates higher detection capability, this flexibility comes at the expense of execution efficiency, generation consistency, and modular structure.
The evaluation also reveals clear limitations. Trigger correctness largely depends on the capabilities of the underlying perception and retrieval components, such as YOLO and CLIP, as well as on the reasoning performance of the large language model and prompt design. Although the framework enforces syntactic correctness and structured execution, it does not mitigate errors originating from these components. Consequently, detection performance primarily reflects the limitations of the integrated models rather than the trigger framework itself.
Future work should investigate combining the modularity, consistency, and runtime efficiency of the DSL-based approach with the flexibility of plain-code generation.
Manually written trigger logic remains advantageous in highly specific scenarios requiring creative problem-solving. However, such implementations lack modularity and scalability, limiting their suitability for evolving deployment environments with dynamic data collection requirements.

\section{Conclusion and Outlook}

This work presented a modular trigger framework for multimodal data collection based on a formally specified domain-specific language (DSL). The DSL enables structured and composable trigger definitions with improved on-device execution on multimodal sensor systems.
Large language models translate high-level user requests into executable DSL configurations. Constraining the generation process through the DSL improves structural consistency while maintaining detection performance comparable to unconstrained code generation. Empirical results demonstrate reduced execution latency compared to plain-code approaches, making the framework suitable for concurrent deployment in resource-constrained systems.
Detection performance remains limited by the underlying perception models, indicating that the primary contribution lies in architectural robustness, modularity, and runtime efficiency rather than improved perception accuracy. 
Future work will focus on increasing expressive flexibility while preserving structural guarantees. Integrating formal DSL constraints with more adaptive reasoning mechanisms may combine the efficiency and consistency of structured generation with the flexibility of unconstrained code synthesis.

\bibliography{references}
\end{document}